\documentstyle[12pt,epsf]{article}
\begin{document}

\baselineskip=24pt

\begin{center}
{\large\bf Temperature dependence of the  band gap shrinkage due to electron-phonon interaction in undoped n-type GaN}
\end{center}

\vspace{0.25in}

\begin{center}
{\bf Niladri Sarkar} and {\bf Subhasis Ghosh} \\
{\sl School of Physical Sciences, Jawaharlal Nehru University, New
Delhi 110067}
\end{center}

\vspace{0.75in}

\begin{center}
{\bf Abstract}
\end{center}

The photoluminescence spectra of band-edge transitions in GaN is studied   as a function of temperature. The parameters that describe the temperature dependence red-shift of the band-edge transition energy and the broadening of emission line are evaluated using different  models.  We find that the semi-empirical relation based on phonon-dispersion related spectral function leads to excellent fit to the experimental data.  The exciton-phonon coupling constants are determined from the analysis of  linewidth broadening.

\newpage

In recent years, GaN and its ternary alloys  have been extensively studied for their application in
blue and ultra-violet light emitting devices, short wavelength
lasers, and electronic devices for high power and high temperature
applications\cite{sn97}. This tremendous application potential
originates from the wide range of direct energy band gaps, the high breakdown
voltage, and the high saturation electron drift velocity.
In spite of technological breakthrough in GaN growth, doping and metal-GaN contacting technologies, several fundamental issues, such as temperature and doping induced band gap shrinkage(BGS) are  still controversial.   In addition to fundamental interest,  these issues are extremely important for the  development and modeling of GaN-based high temperature and high power devices. 
The  temperature induced BGS is observed  in experiments by a monotonic red shift of either band-to-band(BB) and/or excitonic transitions  that are observed in bulk  as well as low dimensional  heterostructures. The  temperature dependence band gap $E_g(T)$ varies from relatively weak  in the low temperature  region to relatively strong at higher temperature region. There is considerable controversy regarding the temperature induced BGS and linewidth broadening.   All the previous studies\cite{ws95,mom96,gdc96,cfl97,kbm04} on GaN have used empirical relations\cite{ypv67,lv84} neglecting the role of phonon dispersion in GaN\cite{jcn98,tr01}to fit the experimental data. There is large variation  of the values of different BGS and linewidth broadening parameters, for example, in case of BGS, the most fundamental parameter   $\alpha$, which is  $-dE_g(T)/dT$  at high temperature limit,  varies  from    0.4meV/K to 1.2meV/K in the literature\cite{ws95,mom96,gdc96,cfl97,kbm04} and in case of linewidth broadening, the most fundamental parameter $\gamma_{LA}$, which is exciton-longitudinal acoustic phonon coupling constant, varies from 0.16meV/K to 0.028meV/K in the literature\cite{gdc96,kbm04}.  Passler\cite{rp99,rp02} has shown the inadequacy of these empirical relations for temperature induced BGS in semiconductors, in particular the inapplicability\cite{rp00} of these relations in case of wide band gap semiconductors. In addition,  residual thermal strain between GaN epilayers and substrates(sapphire and SiC) due to mismatch in lattice constant and thermal conductivity is also responsible for large scattering in the values of different BGS parameters.

In this letter, we present the temperature dependence of BGS in GaN epilayers from 10K to 320K. In particular,  the role of background electron concentration on the different parameters responsible for BGS  and linewidth broadening is studied. The results are compared with different empirical and semi-empirical models for BGS. We have varied the III/V ratio to have excess Ga in the buffer layers prior to the growth of top GaN epilayers.  Previous studies\cite{yk00,ecp99} have shown that excess Ga in the buffer layer improves the crystal quality of GaN epilayers through stain relaxation and increased adatom surface mobility during the initial stage of epitaxial growth. We have recently shown\cite{sd02} that conductivity and the defect-related optical properties of GaN epilayers can be significantly controlled by the III/V ratio in GaN buffer layer. 

In this investigation  we have used  GaN epilayers grown on sapphire
substrates by metal organic chemical vapor deposition(MOCVD).
Prior to the growth of the epilayers, 30nm thick GaN buffer
layer were grown at 565$^o$C. After growing buffer  layer the
substrate temperature was raised to 1000$^o$C for subsequent
1.6$\mu$m thick GaN epilayers.  The background electron concentration in the GaN epilayers were varied by different trimethylgallium(TMG) flow rate in the buffer layer, while keeping the flow rate of  NH$_3$ and H$_2$ constant. The TMG, NH$_3$ and H$_2$ flow rates for all the GaN epilayers were 14$\mu$mol/min, and 1.5 and 2.5slm, respectively.  The electron concentrations were determined by Hall measurements.     We have chosen set of
undoped n-type samples with background electron concentration in the range of 
9.8$\times$10$^{16}$ to 2.7$\times$10$^{18}$ cm$^{-3}$.  The photoluminescence(PL) spectra
were collected in the wavelength region of 340-900nm. The samples
were kept in a closed cycle He refrigerator and were excited with
325nm laser line of  He-Cd laser. The PL signal was collected into
a monochromator and detected with a UV-enhanced Si detector. Fig.1 shows the band-edge transition,  due to  the recombination of the exciton bound to neutral donors(known as I$_2$ exciton)\cite{gdc96b,dv96}, which can be observed at higher temperature($>$300K) due to relatively large  binding energy($\sim$25meV). To avoid the ambiguity in determination of band gap due to  transition from excitonic to  other band-edge transitions at higher temperature, we have limited our experiment till 320K.

Temperature dependence of the PL spectra of  two n-type GaN samples with different background electron concentrations is shown in Fig.1, which shows a maximum redshift of about 65meV as temperature increases from 10K to 320K.   
An appropriate fitting function is required to obtain material-specific parameters from experimentally measured  $E_g(T)$. There are two set of  fitting function for $E_g(T)$ in the literature: (i) empirical  relations proposed by Varshni\cite{ypv67} and Vina et al\cite{lv84}   and (ii) semi-empirical relations based  on  the electron-phonon spectral function $f(\epsilon)$  and the phonon occupation number $n(T)$,  $\epsilon$ is the phonon energy .

The most frequently  used empirical relation  for numerical fittings of E$_g(T)$ was first suggested by Varshini\cite{ypv67} and given by 

\begin{equation}
E_{g}(T)=E_{g}(0)-\frac{\alpha T^{2}}{\beta+T}
\end{equation}

\noindent where E$_g(0)$ is the band gap at 0K, $\alpha$ is the T $\rightarrow \infty$ limiting value of the BGS coefficient $dE_{g}(T)/dT$ and  $\beta$ is a   material specific parameter.   This model represents a  combination of a linear high-temperature dependence with a quadratic low-temperature asymptote for $E_{g}(T)$. Though this phenomenological model  gave reasonable fittings of E$_g$(T)  in elemental,  III-V and II-VI semiconductors with E$_g$ $\leq$ 2.5eV, there have  been several problems with this relation, for example, (i) this relation gives negative values of $\alpha$ and $\beta$ in case of wide band gap semiconductors\cite{ypv67,rp00}, (ii)  $\beta$ is a physically undefinable parameter believed to be related to    Debye temperature $\Theta_{D}$ of the  semiconductor, but this connection has been doubted strongly in several cases\cite{rp02} and    (iii) it has been shown that this relation cannot  describe the experimental data of E$_g$(T) even in GaAs\cite{eg92}. 
 Fig.2(a) shows  the comparison of experimental E$_g$(T) with Varshni's relation\cite{ypv67}. We have obtained the values of $\alpha$ in the range of 0.54 meV/K to 0.63 meV/K and $\beta$ in the range of 700 K to 745K for  the  samples with  electron concentrations of 9.8$\times$10$^{16}$ to 2.7$\times$10$^{18}$ cm$^{-3}$. The parameter $\beta$ in this model only gives an estimation of  ${\Theta_{D}}$, which  is about 870K in GaN. It is clear that though the fitting is good in the low temperature region($<$ 100K), but fitting is poor in the intermediate($\sim$100K) and high temperature region($>$ 200K).

Vina et al\cite{lv84} first emphasized that total BGS $\Delta E_g(T)=E_g(0)-E_g(T)$ is proportional to average phonon occupation numbers $\overline {n}(T)=\left[exp(\frac{\epsilon}{k_BT})-1\right]^{-1}$ and proposed an empirical relation, which can be expressed as 

\begin{equation}
E_{g}(T)=E_{g}(0)-\frac{\alpha_{B}\Theta_{B}}{\exp\left (\frac{\Theta_{B}}{T}\right)-1 }
\end{equation}

\noindent where $\alpha_{B}=\frac{2a_{B}}{\bf{\Theta_{B}}}$, $a_B$ is a material parameters related to electron-phonon interaction,   $\Theta_B=\frac{\hbar\omega_{eff}}{k_B}$ represents  some  effective phonon temperature and $\omega_{eff}$ is the effective phonon frequency. Though this phenomenological model  gave reasonable fittings of E$_g(T)$  in different  semiconductors, there are also  several problems with this relation, for example, (i) at low temperature, this model  shows a plateau behavior of E$_{g}(T)$, which is not observed experimentally, (ii) at higher temperature($\geq$50K), this model predicts    $\Delta E_g(T)\propto exp(-\Theta_B/T)$, but, in most cases  $\Delta E_g(T) \propto T^2$ is observed experimentally,  (iii) it has been shown that this relation cannot describe the experimental data of E$_g(T)$ even in GaAs\cite{eg92}. Fig.2(b) shows the comparison of experimental E$_g(T)$ with Vina's relation.  We have obtained the value of ${\Theta_{B}}$ in the range of 400K to 450K and  $\alpha_{B}$ in the range of 0.40meV/K to 0.44meV/K for  the  samples with  electron concentrations of 9.8$\times$10$^{16}$ to 2.7$\times$10$^{18}$ cm$^{-3}$. There is no doubt that this gives better fitting than Varshini's relation in the high temperature region, but the fitting in the low temperature region is poor.

As mentioned earlier, it can be  shown that the contribution of individual phonon mode to the temperature induced BGS is related to average phonon occupation number $\overline {n}(T)$ and electron-phonon spectral function $f(\epsilon$). Essentially,  $E_g$(T) can be analytically derived from the expression

\begin{equation}
E_{g}(T)=E_{g}(0)-\int d\epsilon {\it f}(\epsilon)\overline{n}(\epsilon,T)
\end{equation}

\noindent The electron-phonon spectral function $f(\epsilon$) is not known {\sl a priori} and extremely complicated to calculate from the first principle. The other option is to use different approximate function for $f(\epsilon)$ to derive temperature dependence of BGS. 
It has been emphasized conclusively\cite{rp99,rp02} that the indispensable prerequisite of estimation of different parameters obtained from the experimentally measured $E_g(T)$ is the application of an analytical model that accounts for phonon energy dispersion. 
The band gap shrinkage  results from the   superposition of contributions made by phonons with largely different energies, beginning from the zero-energy limit for acoustical phonons up to the cut-off energy for the optical phonons. The basic features of phonon dispersion $\delta_{ph}$ and the relative weight of their contributions to  $E_g(T)$, may vary significantly from one material to the other.
The curvature of the nonlinear part of the $E_{g}(T)$ is closely related to the actual position of the center of gravity, $\overline{\epsilon}$($\epsilon=\hbar\omega$), and the effective width  $\Delta\epsilon$, of the relevant spectrum of phonon modes that make substantial contribution to  $E_{g}(T)$ and this has been quantified by phonon-dispersion co-efficient $\delta_{ph}(=\frac{\Delta\epsilon}{\overline{\epsilon}})$.   It has been  shown\cite{rp00}  that  above two empirical models(Eqn.1 and Eqn.3) represent the limiting regimes of either extremely large($\delta_{ph}\approx$1)   in case of Varshini's relation or extremely small($\delta_{ph}\approx$0) in case of Vina's relation. Both these models contradict physical reality  for most semiconductors, whose phonon dispersion co-efficient vary between 0.3 to 0.6\cite{rp00}. 
Several analytical models have been presented using  different forms of the spectral function ${\it f}(\epsilon)$.  Passler\cite{rp99} has proposed the most successful model, which takes  power-law type spectral function,  $f(\epsilon)=\nu\frac{\alpha_{p}}{k_{B}}\left(\frac{\epsilon}{\epsilon_{0}}\right)^{\nu}$
 and  the cut-off energy $\epsilon_{0}$, which  is given by $\epsilon_{0}=\frac{\nu+1}{\nu}k_{B}\Theta_{p}$.
Here, $\nu$ represents an empirical exponent whose magnitude can be estimated by  fitting the experimental data  $E_{g}(T)$.
Inserting, the spectral function ${\it f}(\epsilon)$ and cut-off energy $\epsilon_{0}$ into the general equation for band gap shrinkage(Eqn.2), Passler\cite{rp99} obtained an  analytical expression for $E_{g}(T)$, given by

\begin{equation}
E_{g}(T)=E_{g}(0)-\frac{\alpha_{p} {\bf{\Theta_{p}}}}{2}\left[\sqrt[p]{1+\left(\frac{2T}{\bf{\Theta_{p}}}\right)^{p}}-1\right]
\end{equation}

\noindent where p=$\nu$+1 and  $\alpha_{p}$  is the T$\rightarrow \infty$ limit of the slope dE$_g$(T)/{dT}, ${\Theta_{p}}$ is comparable with the average phonon  temperature\cite{rp99},  ${\Theta_{p}}\approx \overline{\epsilon}/k_{B}$, and the  exponent p is related to the material-specific  phonon dispersion co-efficient $\delta_{ph}$, by the  relation $\delta_{ph}\approx1/\sqrt{p^{2}-1}$. Depending on the value of $\delta_{ph}$,  there are regimes of large and small dispersion which are approximately represented within this model by exponents $p<2$ and $p\geq3.3$ respectively. Fig.2(c) shows the comparison of experimental $E_g(T)$ with Eqn.4.  We obtain an excellent fit to the experimental data.  
 The value of $\alpha_{p}$ obtained here is in the range of 0.50meV/K to 0.54meV/K and the value of ${\Theta_{p}}$  in the range of 500 K to 510 K for  the  samples with  electron concentrations of 9.8$\times$10$^{16}$ to 2.7$\times$10$^{18}$ cm$^{-3}$. The value of the  exponent $p$ which is related to the material specific degree of phonon dispersion is obtained between 2.50 to 2.65 for different GaN samples. The value of $p$ obtained here lies in the intermediate dispersion regime and the  value of the ratio  ${\Theta_{p}}/{\Theta_{D}}$ is 0.58, which means that the center of gravity of the relevant electron-phonon interaction are located within the upperhalf  of the relevant phonon spectra\cite{jcn98,tr01}. Fig.3 shows that the total redshift $\Delta E_g=E_g(10K) - E_g(320K)$ and the phonon dispersion co-efficient $\delta_{ph}$  decrease with electron concentration.   This trend may be explained in terms of the screening of the electron-phonon interaction with higher  electron concentration in the heavily doped samples.

Fig.4 shows the temperature dependence of full width at half maximum(FWHM) of emission PL line in GaN. The FWHM, $\Gamma$(T) can be given by\cite{pl87}

\begin{equation}
\Gamma(T) = \Gamma_0 + \gamma_{LA}T + \frac{\Gamma_{LO}}{\left [ exp\left (\frac{\hbar \gamma_{LO}}{k_BT} \right ) - 1 \right ]}
\end{equation} 

\noindent where $\Gamma_0$  is the intrinsic linewidth at T=0K due to temperature-independent mechanisms, such as defects and electron-electron interaction,  and $\Gamma_{LO}$ is the strength of the exciton-LO-phonon coupling and $\hbar\gamma_{LO}$ is the LO phonon energy. Fig.4 shows the fit to the experimental data using Eqn.5. Insets of Fig.4 shows how $\gamma_{LA}$ and $\Gamma_{LO}$ vary with  electron concentrations.   The estimated LO-phonon energy($\sim$92meV) is very close to the value determined by independent measurements\cite{tr01}. We find exciton-phonon coupling constants are larger than other III-V semiconductors. It has been found that both exciton-LA phonon and exciton-LO phonon interactions become weaker with increasing electron concentration, as shown in Fig.4, signifying the important role of exciton-phonon or electron-phonon interaction on the overall redshift due to temperature induced BGS.   These findings  corroborate with results shown in Fig.3.

In conclusion, we have measured temperature induced band gap reduction and broadening of linewidth in GaN using photoluminescence spectroscopy.   The importance of electron-phonon interaction on the band gap shrinkage has been established.   It has been found that phonon-dispersion based semi-empirical relation is required to explain the experimental data.  Screening of electron-phonon interaction is responsible for decrease of temperature induced BGS in samples with higher electron concentration.

\newpage

\noindent {\bf Figure Captions}

\vspace{0.5in}

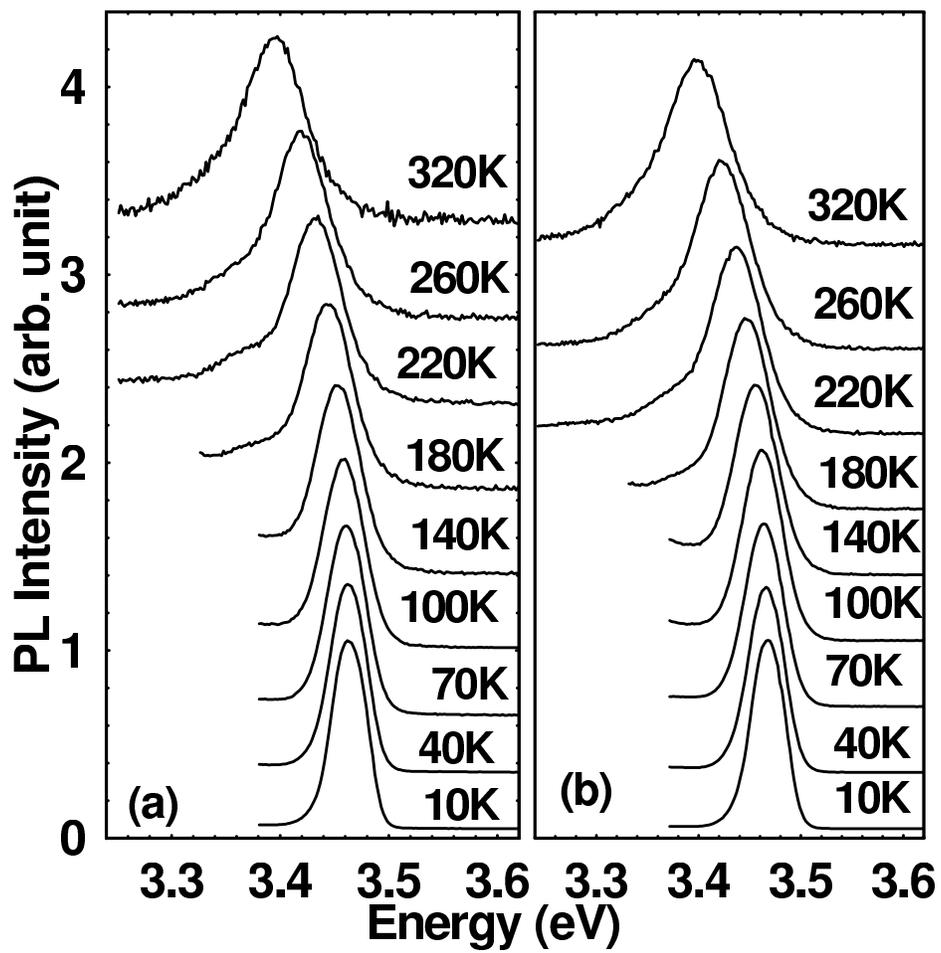
\noindent Figure 1. PL spectra of band-edge excitonic transition at different temperatures from 10K to 320K in GaN epilayers with electron concentration of (a)  1.3$\times$10$^{18}$cm$^{-3}$ and (b) 3.8$\times$10$^{17}$cm$^{-3}$.

\vspace{0.5in}

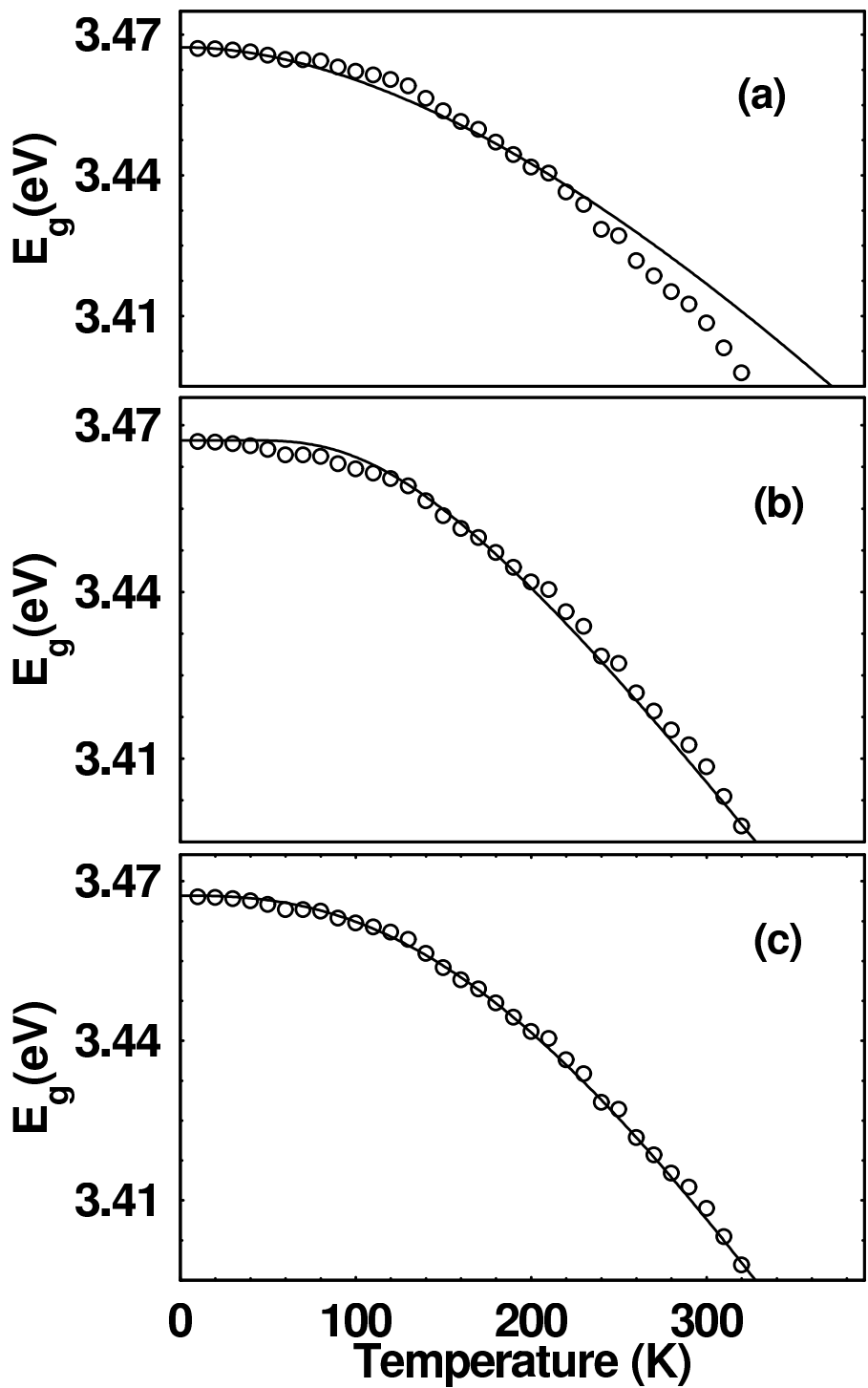
\noindent Figure 2. Temperature dependence of the peak positions of band-edge excitonic transition in GaN sample with electron concentration of  3.8$\times$10$^{17}$cm$^{-3}$. Similar fits have been obtained in case other samples with electron concentration of 9.8$\times$10$^{16}$cm$^{-3}$,  1.3$\times$10$^{18}$cm$^{-3}$ and 2.7$\times$10$^{18}$cm$^{-3}$.   Solid lines are fit to experimental data with (a) Eqn.1, (b) Eqn.3 and (c) Eqn.4.

\vspace{0.5in}

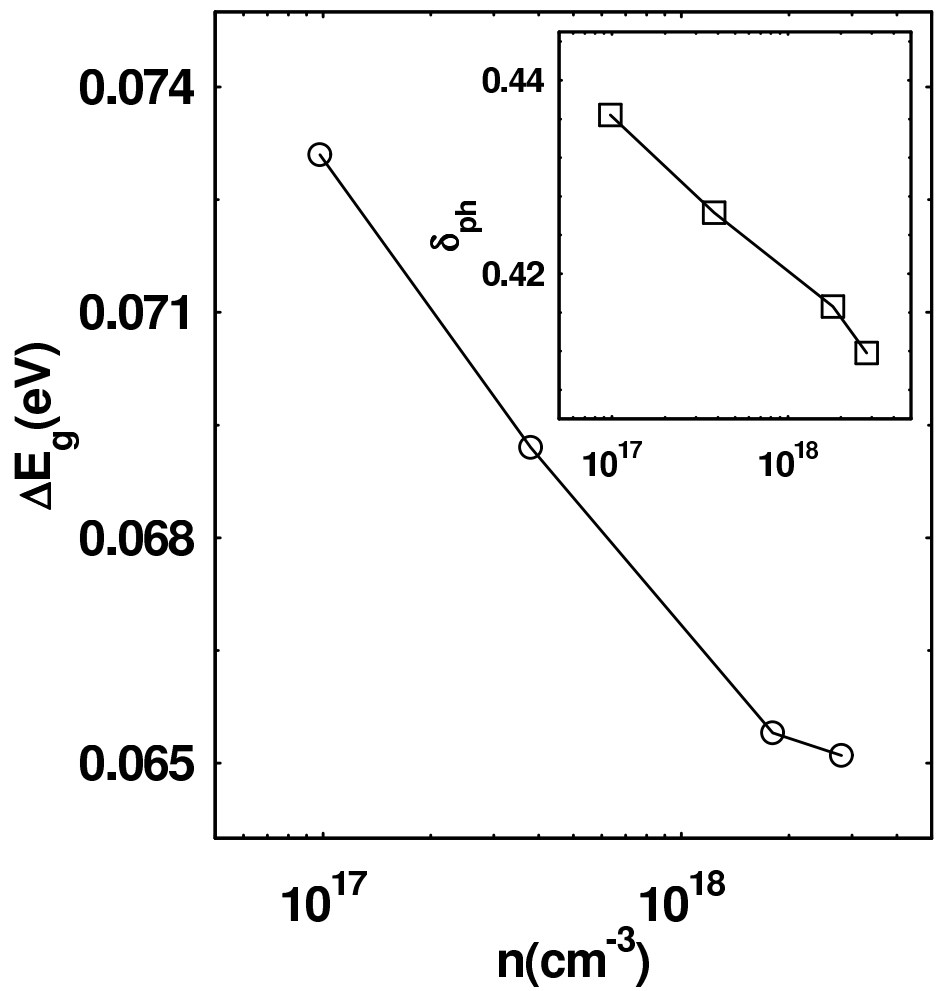
\noindent Figure 3. Electron concentration dependence of BGS, $\Delta E_g=E_g(10K) - E_g(320K)$. Inset shows how the phonon-dispersion parameter $\delta_{ph}$ derived from relation $\delta_{ph} = 1/\sqrt {1-p^2}$  changes with electron concentration $n$. Solid lines are guide for eyes.

\vspace{0.5in}

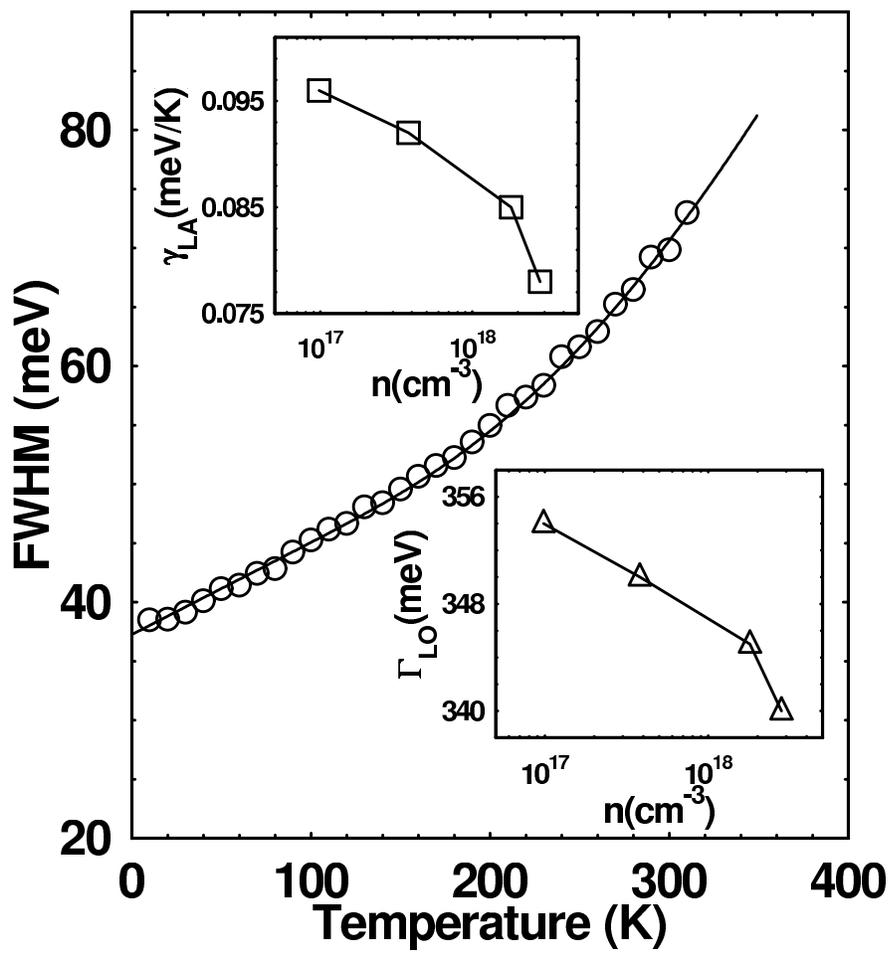
\noindent Figure 4. Temperature dependence of the FWHM of PL emission line for GaN epilayer with electron concentration of 3.8$\times$10$^{17}$cm$^{-3}$. Solid line is the fit to experimental data with Eqn.5. Similar fits have been obtained in case other samples with electron concentration of 9.8$\times$10$^{16}$cm$^{-3}$,  1.3$\times$10$^{18}$cm$^{-3}$ and 2.7$\times$10$^{18}$cm$^{-3}$. Upper inset shows how electron-acoustic phonon coupling constant($\gamma_{LA}$) varies with electron concentration $n$. Lower inset shows how electron-optical  phonon coupling constant($\Gamma_{LO}$) varies with electron concentration $n$. Solid lines are guide for eyes.

\end{document}